\documentclass[fleqn,10pt,twocolumn]{wlscirep}

\usepackage{amsmath}
\usepackage{amssymb}
\usepackage{color}
\usepackage{graphicx}
\usepackage{braket}
\usepackage{textcomp}

\begin{document}
\title{Relativistic quantum key distribution system with one-way quantum communication}
\author[1,2,*]{K. S. Kravtsov}
\author[1,2]{I. V. Radchenko}
\author[1]{S. P. Kulik}
\author[3,4]{S. N. Molotkov}

\affil[1]{Faculty of Physics, Moscow State University, Moscow, Russia}
\affil[2]{A.M. Prokhorov General Physics Institute RAS, Moscow, Russia}
\affil[3]{Academy of Cryptography, Moscow, Russia; Institute of Solid State Physics, Chernogolovka, Moscow Rgn., Russia}
\affil[4]{Faculty of Computational Mathematics and Cybernetics, Moscow State University, Moscow Russia}
\affil[*]{ks.kravtsov@gmail.com}

\date{\today}
\begin{abstract}
Unambiguous state discrimination (USD) is one of the major obstacles for {\em practical} quantum key distribution (QKD). Often overlooked, it
allows efficient eavesdropping in majority of practical systems, provided the overall channel loss is above a certain
threshold. Thus, to remain secure all such systems must not only monitor the actual loss, but also possess a comprehensive
information on the safe {\em 'loss~vs.~BER'} levels, which is often well beyond currently known security analyses. The more
advanced the protocol the tougher it becomes to find and prove corresponding bounds. To get out of this vicious
circle and solve the problem outright, we demonstrate a so called {\em relativistic} QKD system, which uses causality to
become inherently immune to
USD-based attacks. The system proves to be practical in metropolitan line-of-sight arrangements. At the same time it has
a very basic structure that allows for a straightforward and comprehensive security analysis.
\end{abstract}

\maketitle

\section*{Introduction}
The workhorse of quantum cryptography has always been the BB84~\cite{BB84} protocol, whose elegance mainly comes from the use of true single
photons. It also enables quite unique and comprehensive security proof~\cite{SP00,CRE04,TLG12}. 
At the same time, no practical QKD protocol can use the same information carriers: they have to rely upon weak coherent pulses (WCPs) instead.
As WCPs are formally infinite-dimensional quantum systems, there is always a non-zero probability of unambiguous
discrimination of the transmitted states in the channel~\cite{CB98,AC00,DJL00}. Thus, starting from some level of loss,
conventional WCP-based
QKD systems inevitably lose their guaranteed security. Corresponding thresholds are well-known for simple protocols as
B92~\cite{P88} and WCP-based BB84~\cite{DJL00}, but, to the best of our knowledge, still far from being found for
popular WCP-based COW~\cite{SBG05} and
DPS~\cite{IWY03} protocols, whose security proofs, thus, may be considered incomplete. To avoid this potential security breach at
high channel loss we argue that additional measures have to be taken in the protocol design to
completely disallow masking unsuccessful unambiguous state discrimination (USD) in losses.

Earlier, many efforts were directed towards developing protection against a much more narrow than USD type of attack, the
photon number splitting (PNS). They resulted in various decoy state strategies~\cite{H03,W05}, which seem to help providing protection
against PNS by the cost of monitoring multiple additional channel statistics besides the simple loss. To the best of
our understanding, these strategies lack simple security grounds and fail to show clear and universal security proof.
In general, this approach just admits the faulty use of WCPs in BB84-like protocols and tries to make up additional
measures to save these (impractical) protocols, instead of finding a universal solution.

A better idea is to design new protocols, where WCP nature of information carriers is already accounted for.
A valid approach known from the early days is the B92 with a strong phase reference~\cite{B92}, where the presence of the strong
reference pulse makes it impossible for Eve to send vacuum states if the USD fails to get a conclusive result. Another
alternative first coined in~\cite{GV95} with single photons and later re-invented in a practical form in~\cite{RKK14} is to use
relativistic limitations. They allow to force Eve make decisions about her actions {\em before} she can actually measure
the state in the line, thus breaking her only winning strategy due to causality. In this
paper we demonstrate an improved experimental realization of this protocol, where we implemented an efficient
one-way configuration with an active single-mode free-space channel tracking system, demonstrating stable operation
over 180 m.

\begin{figure}
\centering
\includegraphics[width=\columnwidth]{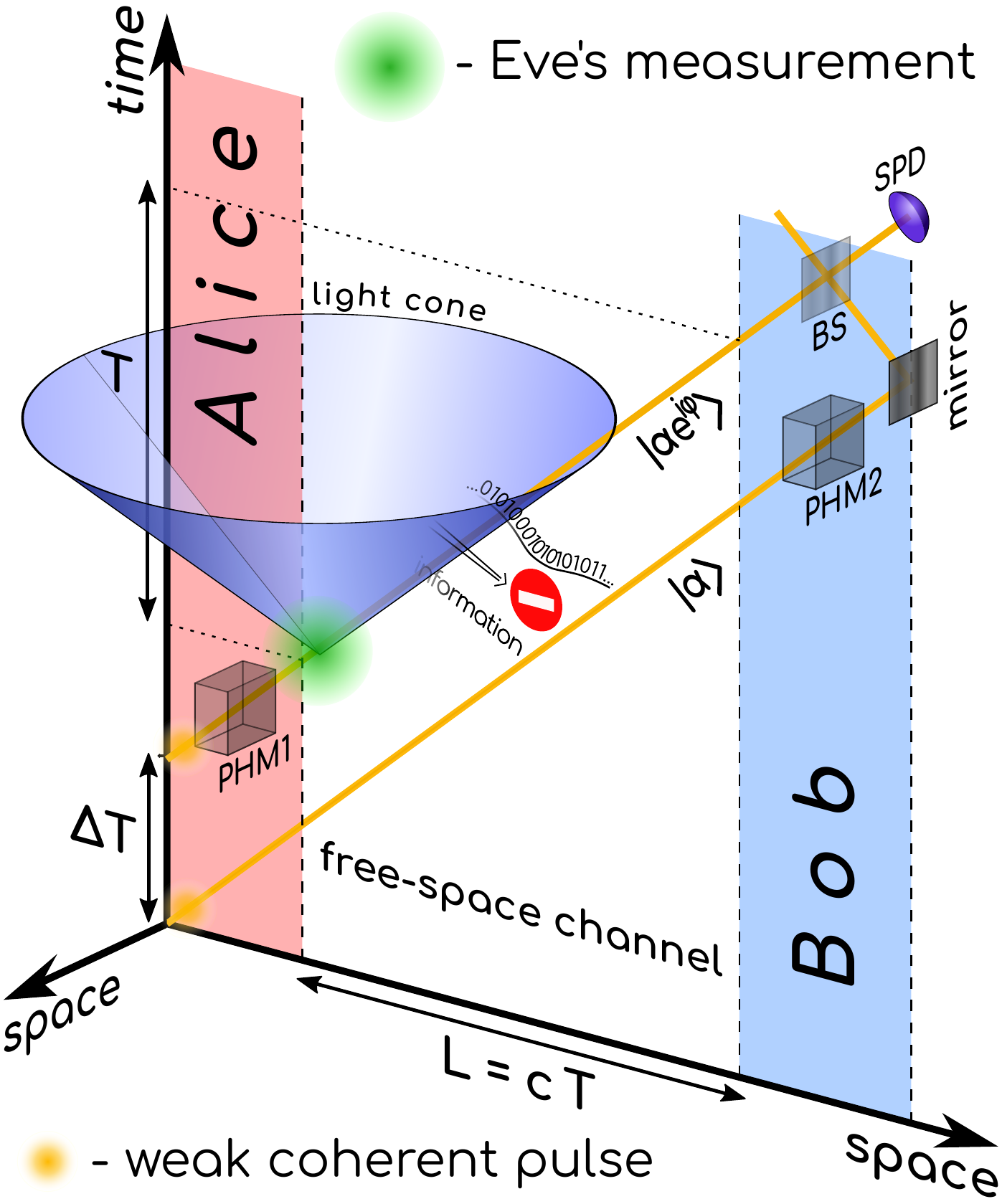}
{\caption{Space-time diagram of the relativistic protocol. The two pulses travel with the speed of light, thus,
forbidding Eve's actions on the first pulse dependent on her measurement of the second, modulated one. PHM -- phase
modulator, BS -- symmetric beamsplitter, SPD -- single-photon detector.}\label{fig_proto}}
\end{figure}

\section*{Results}

The relativistic protocol is schematically shown in Fig.~\ref{fig_proto} as a space-time diagram. Its key component is
the quantum transmission with the speed of light in two time windows separated by a measurable time interval $\Delta T$. The key
generation procedure looks very much like B92 protocol~\cite{B92}. To establish one bit of the raw key Alice and Bob
randomly choose one bit of information each: $b_A$ and $b_B$ respectively, where $b \in \{0,1\}.$ Alice transmits two pulses:
a reference WCP $\ket{\alpha}$ in the first time window and a signal $\ket{e^{ib_A\varphi}\alpha}$ in the second. Bob applies a
phase shift of $b_B\varphi$ to the first time window and measures the result of interference between the two. He can only detect a
photon if $b_A \ne b_B$, otherwise there is a destructive interference between the two pulses and, therefore, the vacuum
state in the detector. So any time Bob's detector
clicks, he tells this to Alice and they end up with one more bit of the raw key.

In the conventional B92, eavesdropping strategy is straightforward. There is a certain probability of USD between
$\ket{\alpha}$ and $\ket{e^{i\varphi}\alpha}$, so whenever Eve succeeds in her measurement she retransmits the correct state.
If the USD fails Eve blocks both pulses, so the overall effect is indistinguishable from the genuine lossy channel.

The {\em relativistic} protocol ensures that at any moment the first pulse lies outside the light cone generated by the
second one,
so there cannot be any causal connection from the second to the first one. Therefore, any Eve's measurements of the
data pulse cannot affect her actions on the reference.
To ensure the proper space-time relation between the pulses, the distance $L$ between Alice
and Bob should be known a-priori as it is a critical security parameter of the protocol. All signals delayed in the
channel by more than $L/c$, where $c$ is the speed of light, are ignored. 
In this modified framework Eve has no ability to block the reference pulse depending on her measurement result, as it
would contradict the causality principles. However, if any one of the two pulses in the channel is missing, Bob sees the results uncorrelated with the
states Alice sent, producing errors in the raw key. So whenever Eve lets the reference through, but {\em then} fails to
measure the data pulse, she causes errors in the raw key. On the contrary, if she blocks the reference, but retransmits
even the correct data pulse, she causes errors too.
This picture has much in common with the strong phase reverence version of B92. The strong classical reference cannot be
removed from the channel because this is directly detectable. If, on the contrary, it is present without the correct WCP
companion, it inevitably produces errors. While both approaches offer ultimate protection against USD, we believe that
our relativistic protocol is less technology demanding and can be more practical.

Each detector click gives Bob one bit of information, as he
effectively performs post selection, i.e. chooses only those pulses for which his measurement succeeds. On the contrary,
Eve's actions cannot depend on her measurement results, otherwise
Bob would see uncorrelated with Alice bits. When Eve gets a measurement result, i.e. not earlier than the second time
window, it is already too late to reach the first window, located beyond the light cone~\cite{M12}. Therefore, her information per
any channel use is fundamentally limited by the capacity of such binary quantum channel, i.e. by the Holevo quantity~\cite{H98}.
The difference between Bob's information and the fundamentally limited information of Eve gives the room for secret key
generation. That is, transmission with the speed of light over the known distance, together with precise synchronization and timing, gives
a new security component to QKD that offers assured protection against USD-based and any intercept and resend attacks.

An efficient experimental realization is another question addressed in this work.
Transitioning to a one-way quantum channel configuration makes the system more protected from Eve's actions, compared to the double-pass
one~\cite{RKK14}, where Eve could manipulate classical pulses traveling from Bob to Alice. It also greatly improved the operation
rate of the system, as there is no need to wait the round-trip time to send more data into the channel.

As the security of such causality-based {\em relativistic} protocol relies on precise timing, synchronization plays a critical role in
the protocol. Malicious altering of the synchronization process may easily break the foundations of the protocol
security, opening a backdoor for eavesdropping. Therefore a special secure procedure was
developed to guarantee proper synchronization during the protocol operation. It requires a backward classical channel
where information travels with the speed of light.

To initiate quantum transmission Bob generates a random bit sequence and sends it to Alice using the
classical channel with the same rate that Alice uses for QKD. At each received bit Alice stores it in her local memory and transmits
one WCP into the quantum channel. After the whole packet is
transmitted, Bob and Alice compare their synchronization sequences. If the sequences are the same, Alice can
guarantee that she received each bit not earlier than Bob expects her to get it. Otherwise, it would be a superluminal
information transfer between Bob and Alice, which contradicts the relativity theory. That directly means that Alice never sent any quantum
state into the channel earlier than Bob thinks she did. At the same time, this is the only case when Eve would have an extra time to
act {\em after} her measurement without causing errors: if she could force Alice to transmit earlier than Bob thinks,
the protocol would be broken. If, on the contrary, Alice sends her pulses later, Bob just will not receive any correlated with
Alice raw key, so the packet will be discarded as not containing any secret information.
If after comparison the synchronization sequence received by Alice appears to differ from that of Bob, it
is a potential sign of an ongoing synchronization attack and the whole packet must be discarded as unreliable.

The backward communication channel required for synchronization is realized via the tracking system, which also serves
for transmission of service data and control messages in both directions between the parties.
Besides data communication, the tracking system is needed to keep the quantum channel up, as, in the contrast to
conventional free-space QKD systems, the present one needs a single mode receiver, which is compatible with a fiber-based delay
interferometer. Without active tracking, the system was extremely unstable when mounted on standard theodolite tripods
and would not operate reliably even for a few minutes. With the tracking system implemented it showed good performance
at least for hours, although we did not check the stability for a longer time. More detailed information about the
single mode channel and the tracking system can be found in Supplementary. There is also a discussion about the
difference between the group velocity of pulses in the air and the speed of light, which is insignificant for the
implemented parameters of the protocol.

\begin{figure*}
\centering
\includegraphics[width=1.8\columnwidth]{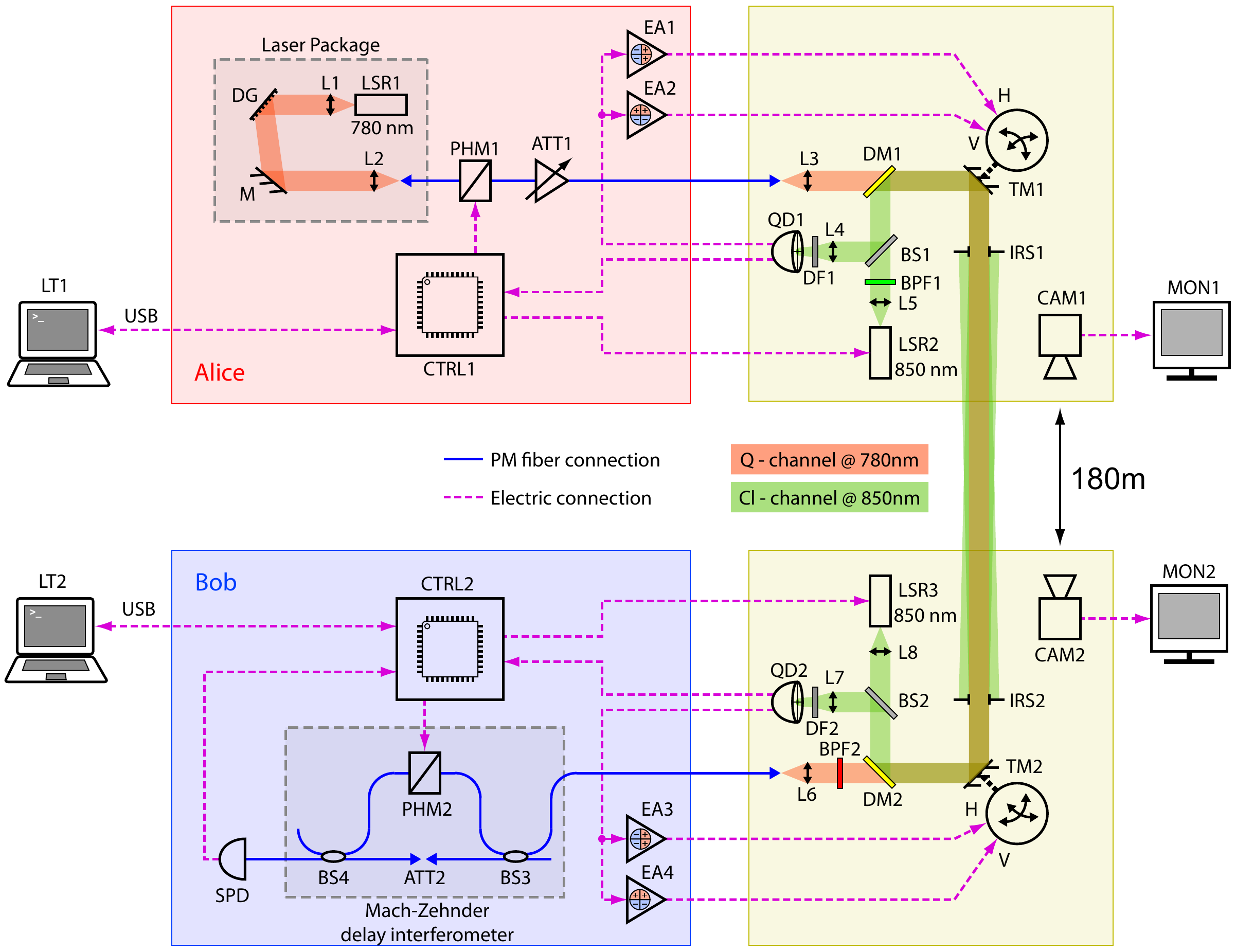}
{\caption{Experimental setup schematic. LT~--~laptop-based user station; DG~--~diffraction grating in Littrow
configuration; L~--~lens; M~--~mirror; PHM~--~150~MHz lithium niobate fiber-coupled phase modulator; ATT~--~variable
optical attenuator; CTRL~--~control electronics; EA~--~electronic error amplifier in the tracking feedback loop;
DM~--~dichroic mirror; TM~--~piezo tip-tilt mirror; QD~--~quadrant photodetector; DF~--~ground glass-based diffuser;
BS~--~symmetric beamsplitter; IRS~--~25~mm iris diaphragm; BPF~--~band-pass filter; CAM~--~coarse pointing camera;
MON~--~user monitor for the camera; SPD~--~silicon avalanche photodiode-based single-photon detector.}\label{fig_setup}}
\end{figure*}

Another experimental challenge addressed in our one-way design is the proper alignment of the receiving side interferometer.
To simplify the setup we eliminated the transmission side delay interferometer altogether and used a CW laser instead.
Thus, Alice's side contains only a narrow linewidth CW laser (external cavity diode laser), a phase
modulator and an attenuator, as shown in Fig.~\ref{fig_setup}. The receiving side has a polarization maintaining fiber
based delay interferometer with a phase modulator in one of the arms, which serves for both interferometer alignment
(with a quasi-DC bias) and data modulation during the QKD stage. The bias is constantly adjusted according to  the number of single
photon detector clicks when biased at $\pi/2$ below and above the normal level that corresponds to the dark interferometer
output. A whole cycle of the modulator work is shown in Fig.~\ref{fig_timing}. More details on interferometer alignment
are found in Supplementary.

\begin{figure}
\includegraphics[width=\columnwidth]{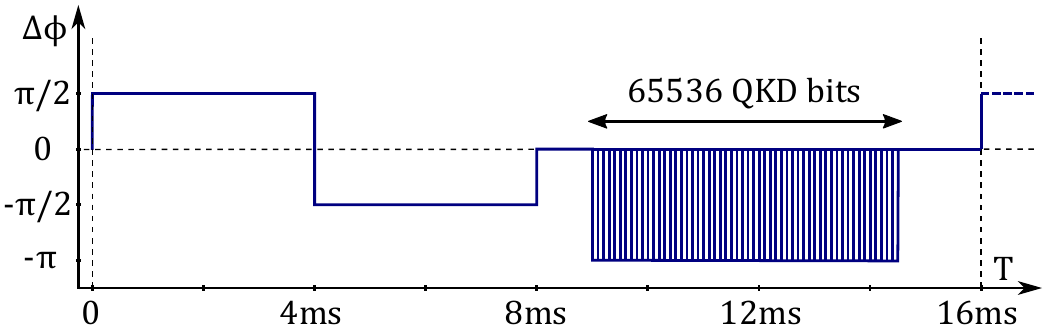}
{\caption{Operation of the receiving interferometer phase modulator. In each 16~ms cycle first it measures count frequencies in
two quadrature points to adjust the bias, and then proceeds to the QKD sequence.}\label{fig_timing}}
\end{figure}

The main operation parameters are as follows. Each transmitted quantum symbol is a 10~ns long piece of the CW laser signal at
$\lambda=780$~nm with the output intensity of -92.9~\dots-78.9~dBm, which corresponds to 0.02~\dots0.5~photons per
pulse. The delay $\Delta T$ in the receiving interferometer is 20~ns, so each symbol interferes with the corresponding
chunk of the CW signal going $\Delta T$ ahead (the phase reference window).
The depth of phase modulation equals $0.8\pi$. Phase modulated symbols come in packets of 65536 bits each with the average rate of
25~MHz.
A packet can be sent in any phase modulator cycle, which is 16~ms long (see Fig~\ref{fig_timing}). However, the actual packet rate was limited by
the time needed to exchange the random data buffers and measurement results with a PC via a USB interface, so the actual
rate was about 2~packets/sec. 

\begin{figure}
\fbox{\includegraphics[width=\columnwidth]{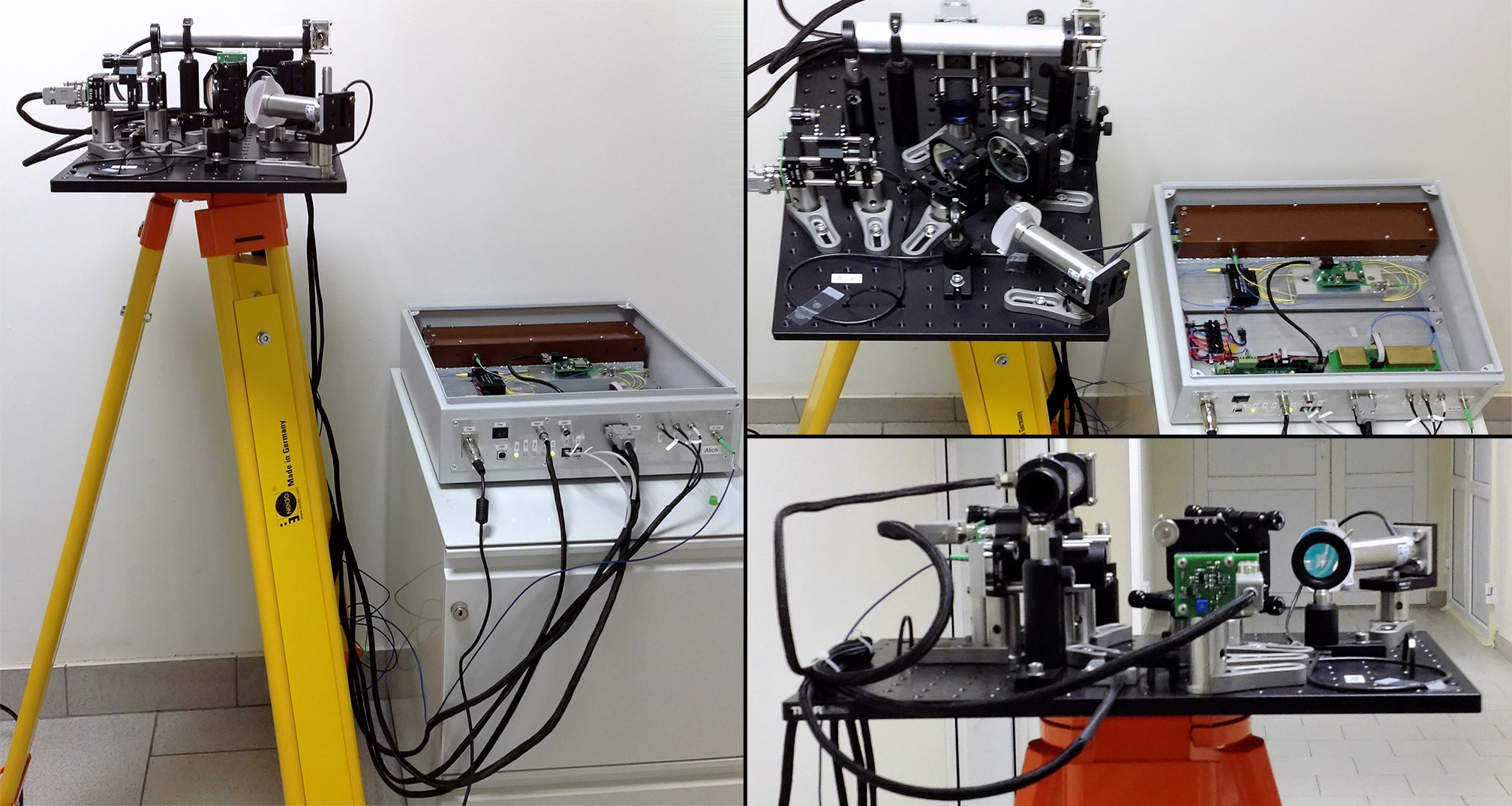}}
{\caption{Station Alice: a tripod with a free-space channel tracking platform and a box with fiber optic components and
all electronics. }\label{fig_photo}}
\end{figure}

The whole system consists of two similar stations, each containing a box with electronics and fiber-based elements, and a free-space
channel tracking platform placed on a tripod as shown in Fig.~\ref{fig_photo}. The quantum single-mode free-space channel uses
diffraction-limited 1"~diameter aspheric lenses to collimate radiation to/from single-mode polarization maintaining
fibers. Quantum signals are spatially mixed with the 850~nm beacon radiation used by the tracking system.
Beacon light is
detected by a quadrant photodiode to provide feedback to the piezo driven steering mirror. It also delivers a 25~Mbit/s
Manchester encoded classical signal used for secure synchronization and transfer of auxiliary information between stations.
The tested channel length of 180~m was actually limited by the length of the building, while the system itself was designed to operate
over as far as 400~m.

\setlength{\fboxsep}{1pt}

\begin{figure}
\includegraphics[width=\columnwidth]{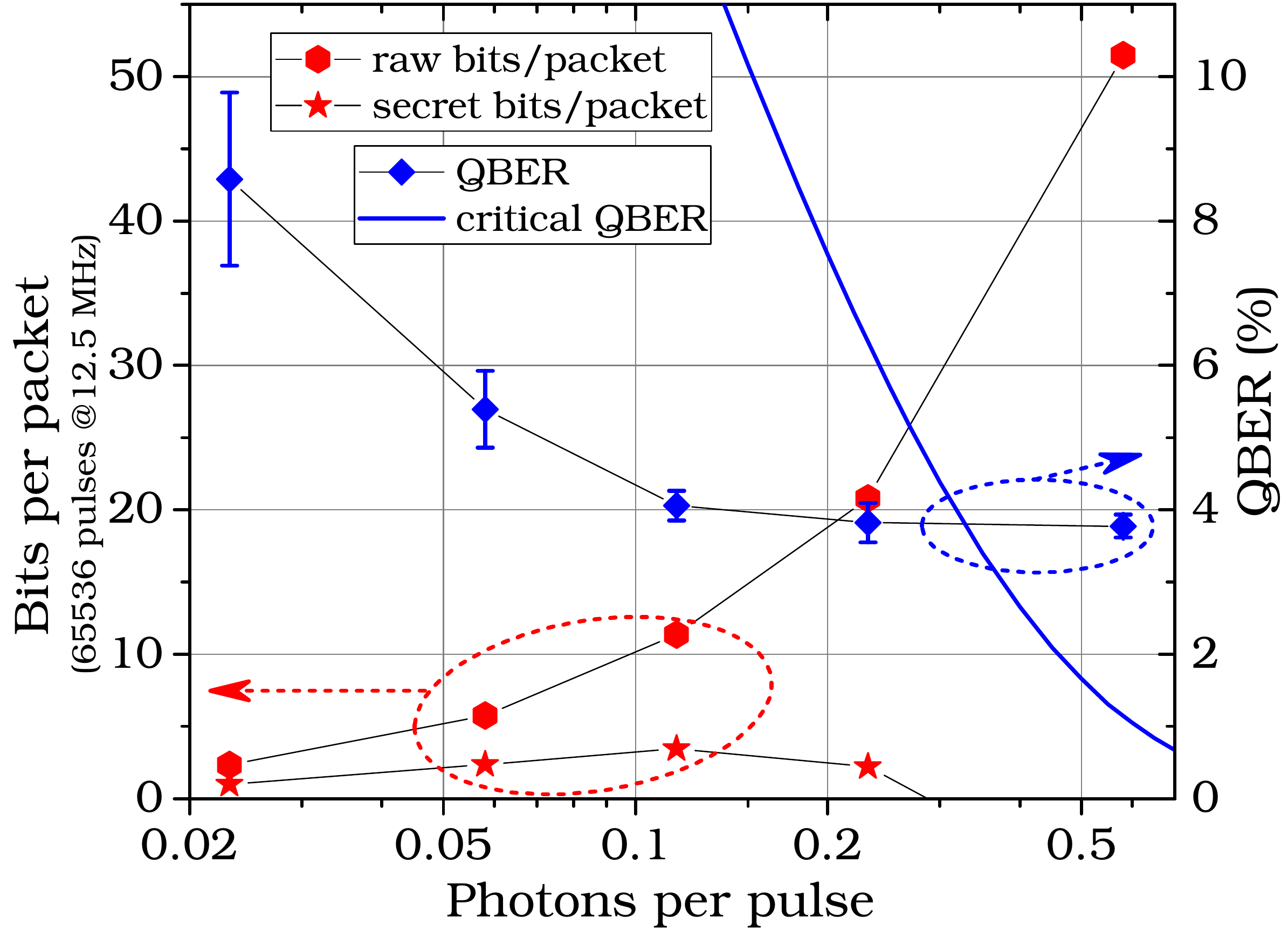}
{\caption{System efficiency and QBER measured in PRBS operation mode vs. the average photon number. The figure also
shows the calculated number of
asymptotic secret bits per packet as well as the critical QBER, above which no secret bits can be extracted.
Error bars on the QBER plot are purely statistical ones corresponding to the uncertainty of QBER estimation based on the
finite number of obtained bits. More precisely, they depict a 95\% binomial proportion confidence interval for all
raw bits accumulated in a particular setting. }\label{fig_prbs}}
\end{figure}

The system operates in two modes: with pseudo-random bit sequences (PRBS) and with real random data. The first one is used
for testing purposes as it provides an easy way to calculate quantum bit error ratio (QBER) without utilization of the
classical channel 
(stations know the pseudo-random sequences used at the other end of the line). The second mode works with real random data
from a quantum random number generator (QRNG)~\cite{KRK15} stored at laptops. Figure~\ref{fig_prbs} shows system efficiency and
QBER for the PRBS operation mode at different photon levels.

To estimate the asymptotic secret key rate we use the information based approach. The raw information obtained by Bob
must be reduced to eliminate the Eve's information, or more accurately, the information that could potentially leak to
Eve. As the raw key always contains some errors, a portion of the raw key is also used for error correction.
As discussed earlier, the implemented relativistic scheme disallow Eve's influence on the received quanta in the way that her actions depend on
results of her measurements. Without this ability to post-select, Eve cannot decide which pulses will travel to Bob
and produce detector clicks and which she will block contributing to the channel loss. At most she can obtain the
average information per pulse. Effectively, Eve's information is bounded by the Holevo quantity~\cite{H98}:
$$\chi(\mu,\varphi) = h\left(\frac{1-\exp(-2\mu\sin^2(\varphi/2)}2\right),$$
where $h(p)=- p\log(p) - (1-p)\log(1-p);$ $\varphi = 0.8\pi$ is the modulation depth, and $\mu$ is the average number of
photons per pulse.
Ideal asymptotic error correction requires $h(\textrm{QBER})$ bits, so
the overall asymptotic secret key rate equals $\mathbf{R} = 1-\chi(\mu,\varphi)-h(\textrm{QBER}).$
It should be noted that here we do not take into account any finite-size effects, as they do not qualitatively change
the results. Some elaboration for finite-sized sequences is published elsewhere~\cite{M12}.

At small $\mu$ Eve's information is small, but the estimated secret key length is severely limited by the high QBER. At
large $\mu$ QBER decreases, however, the Eve's information becomes the limiting factor. The maximum efficiency is
observed at around $\mu = 0.1$ as follows from Fig.~\ref{fig_prbs}.

\begin{figure}
\includegraphics[width=\columnwidth]{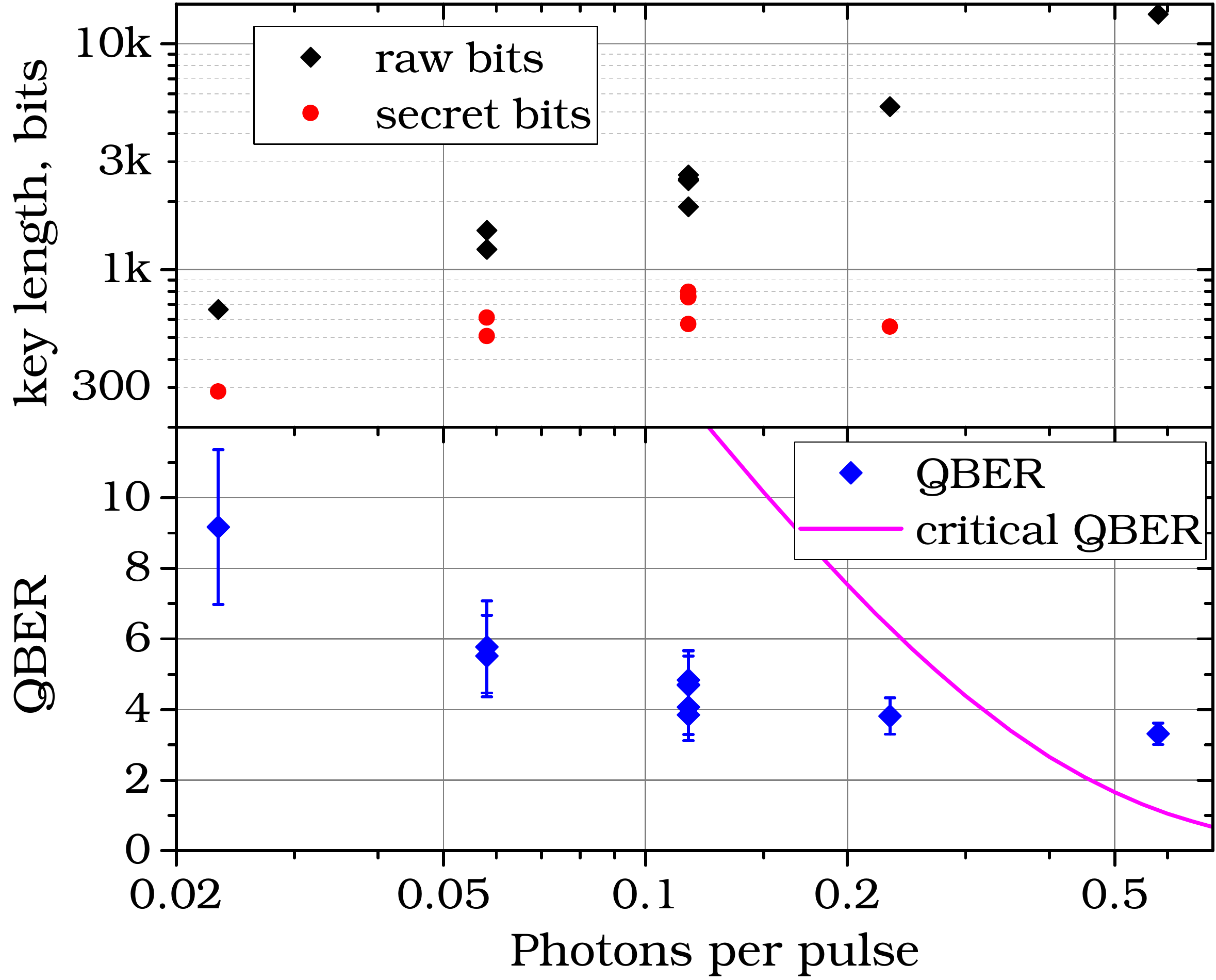}
{\caption{Key lengths and QBERs vs. the average photon number for QKD with random data from the QRNG. Each point is a result of
QKD with 16~Mbit input buffers, i.e. 256 packets transmitted. Error bars on the QBER plot show the 95\% binomial proportion
confidence interval for the particular raw key obtained in the corresponding data point.}\label{fig_qrng}}
\end{figure}

Operation with real data from QRNG was performed to distribute actual raw keys. Privacy amplification and error
correction was not implemented in the experiment, as it is relatively straightforward, but too time consuming for this
proof of principle demonstration. Therefore, all estimations are made using the asymptotic relation found above and the
obtained raw keys. Figure~\ref{fig_qrng} shows
experimentally measured data --- raw key length and QBER --- as well as asymptotically estimated number of secret bits.
Each data point shows the result of a particular exchange of $1.68\times 10^7$ WCPs between Alice and Bob.
For some photon numbers per pulse we made a few measurements to ensure the repeatability of the results, for other just
a single key exchange was performed.
The most efficient secret key generation was observed at $\mu = 0.116$, where the raw key generation rate (inside a
packet) equals 2170~bits/sec and the asymptotic secret key rate is estimated as 660~bits/sec. As mentioned earlier,
average rates are substantially smaller due to the slow data exchange with laptops: 20 and 6.2 bits/sec
respectively.

The performance of the single-mode free-space channel is another thing to mention. Although it was inside a building,
heating and ventilation caused a significant mode distortion. Typical wandering frequency is measured to be below
10~Hz, so the tracking system with the bandwidth of 10's of Hz substantially helped in
reducing the loss. Nevertheless, active tracking could only compensate for the beam shift as a whole, but not the
mode distortion. The measured free-space quantum channel loss (the ratio between the transmitted power and the Rx
fiber-coupled power) is
around 13~dB. At the same time the overall system efficiency, i.e. the ratio of detected photons to the transmitted
ones, was $1.5\times10^{-3}.$

\section*{Discussion}

The presented concept of the {\em relativistic} or causality-based QKD provides a new dimension to conventional quantum
cryptography. Its main advantage is in complete decoupling between the channel loss and the security level. No
additional tests are required (at least in theory), besides the standard privacy amplification and error correction, to
guarantee information theoretic key security. In this sense it has much in common with the original B92 protocol with
strong reference pulses. At the same time, the presented protocol seems to be less technology demanding, as, to the best
of our knowledge, there is no experimental demonstration of the original B92 yet. This comes in exchange for the
additional assumptions that we need from the channel, namely, the knowledge of the channel length.

The channel length or, more precisely, {\em distance} between Alice and Bob plays a critical security role in the
relativistic protocol. It is an important security parameter, which should be known a-priori to guarantee the protocol security.
Formally, one cannot be more confident in the security of the generated keys, than he is confident in the distance
between the parties. However, this can be eased by placing a restriction only on the lower bound of the channel length.

In fact, increasing the delay $\Delta T$ between two pulses one can tolerate more deviation between the actual time of
flight and $L/c$. There is a more detailed discussion on that subject in Supplementary, but in general one has to
make sure that the second pulse cannot overtake the first one even if the second one travels along the straight line between Alice
and Bob with the speed of light. Thus, the minimal required delay between the pulses equals $\Delta T_\mathrm{min} =
2(T_o - L_\mathrm{min}/c ),$ where $T_o$ is the {\em observed} time of flight, $L_\mathrm{min}$ is the lower confidence bound for
the value of $L$, and the factor of 2 is included because in this particular implementation the synchronization process
relies upon the same channel and therefore can be offset by the same amount. It could be, however, cut in half if an
external trusted synchronization scheme is used.

As $L$ is always positive, $\Delta T > 2T_o$ is a safe, but often impractical choice. To remain practical, one would want
$\Delta T \ll T_o$. This is feasible for a large-distance free space communication with a moving target confined in
some relatively small area, e.g. inside a town. Another possible strategy is using hollow core photonic crystal fibers
(PCFs), where the effective refraction index is demonstrated to be as low as 1.003 and the optical loss is expected to
beat that of conventional silica fibers~\cite{PWP13}. Future hollow core PCF infrastructure may become the natural
backbone for the relativistic QKD network, since the difference between the propagation speed and $c$ is minimal in
such fibers.

Yet another practical possibility is to use the same phase-encoding hardware either in conventional (when no reliable
information about distance is available) or relativistic mode. This may be a good compromise for attaining the best
possible security scenario depending on the particular circumstances.

In conclusion, we report a {\em relativistic} QKD system, which, unlike conventional protocols, offers inherent
resistance against USD based attacks under arbitrary large channel loss while using practical weak coherent pulses as information
carriers. Our experimental setup operates via a 180~m uni-directional single-mode free-space quantum channel with the
active tracking system. Due to its simple structure and straightforward security foundations, this protocol may become
the first {\em practical} QKD protocol with as general a security proof as for BB84. Its advantages are best attained in
line-of-sight metropolitan links up to several kilometers long between stationary objects or in the future low loss
hollow core PCF networks, where ultimate security needs
meet the ease of experimental realization.

\section*{Methods}

\paragraph{Hardware implementation.} The light source is a CW-driven 90~mW 780~nm laser diode with an external cavity based on a
1800/mm diffraction grating in Littrow configuration. Phase modulators used are low frequency PM fiber coupled lithium niobate 
modulators that have no internal electrical waveguide and termination. Unlike traveling wave modulators, this type can be used for simultaneous
interferometer adjustment and high-speed phase modulation due to their tolerance to large DC offsets. The single photon
detector is based on a silicon Geiger mode avalanche photodiode package with the internal thermoelectric cooler. Its
quantum efficiency is 35\% and the dark count rate is around 700~Hz. The tracking system uses PI S-330.80L piezo tip-tilt
platforms with 2'' mirrors. The main resonance frequency for this steering mirror configuration is around 920~Hz.
As a beacon light source we use a 10~mW directly modulated 850~nm laser diode. Its radiation is collimated using 0.5NA
$F=8$~mm aspheric lens.
Quadrant photodiodes have $3\times3$~mm$^2$ active area and they are placed in the focal plane of the $F=80$~mm focusing
lens. To smooth the feedback response a 1500~grit ground glass diffuser is placed a few mm before the photodiode.
The AC component of the detected signal is summed from all the quadrants, is frequency corrected, amplified and
converted to a binary data stream  --- the classical communication channel. The DC component is amplified separately
for all the quadrants and then the vertical and horizontal error channels are formed by pairwise subtraction of
corresponding signals. The error signals are scaled with respect to the total received power and are input into the two PID
control loops. Fine synchronization between the stations is performed by the PLL which locks to the received digital
waveform of the classical channel. The used Manchester encoding ensures that there is enough zero crossings for the PLL to
operate regardless of the transmitted data.

\bibliography{refs}

\section*{Acknowledgements}
This work was partially supported by the RFBR grant No. 17-02-00966 and Russian Ministry of Education and Science grant
No. 03.625.31.0254.

\section*{Supplementary materials}

\subsection*{System operation and data flow patterns}

\begin{figure*}[h!]
\centering
\includegraphics[width=1.4\columnwidth]{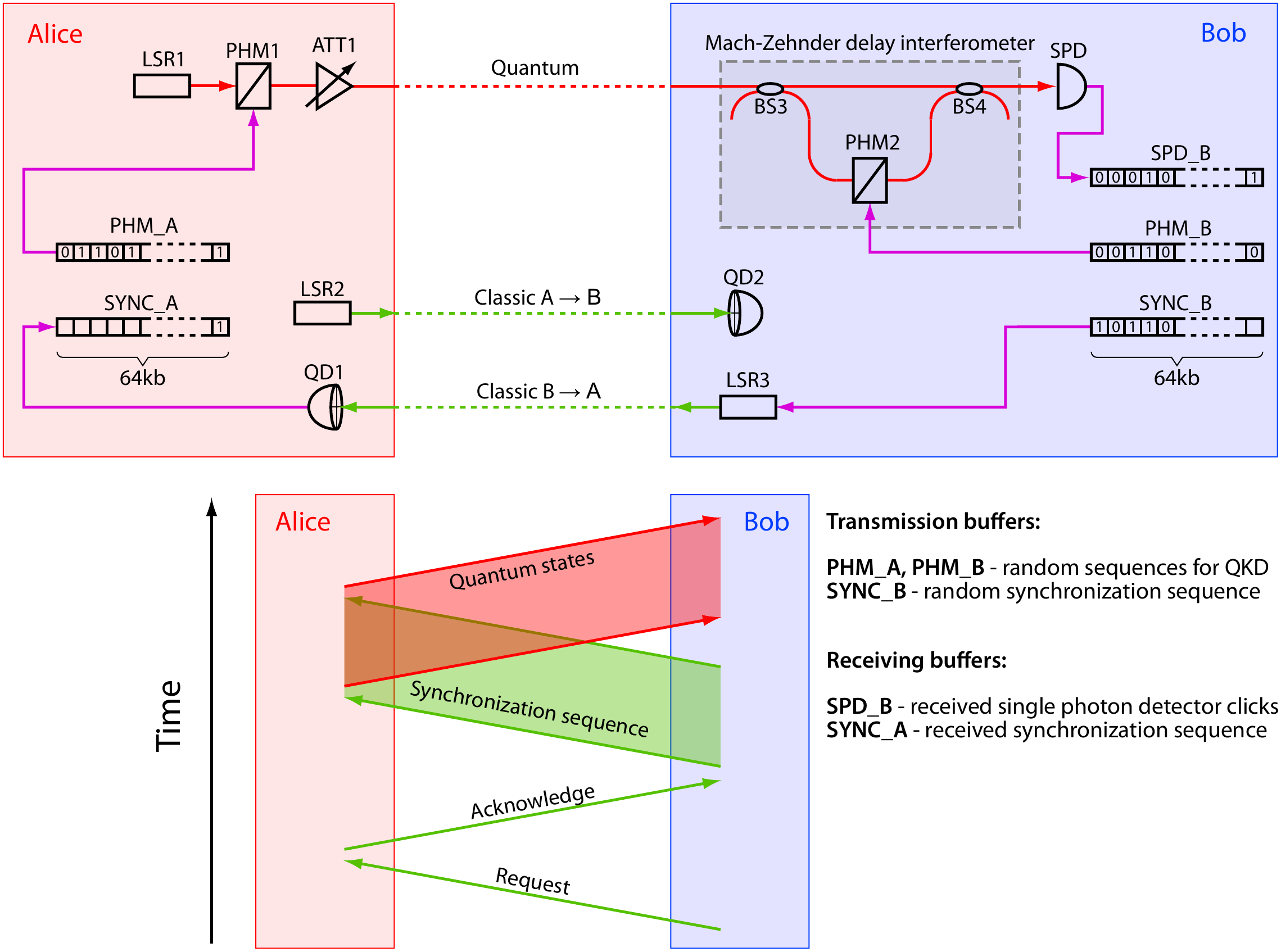}
{\caption{Data flow chart during system operation.}\label{fig_flowchart}}
\end{figure*}

Figure~\ref{fig_flowchart} shows the data flow chart within the system. Before the protocol starts, Alice and Bob obtain
random sequences from the laptops and store them in buffers PHM\_A, PHM\_B, and SYNC\_B. When both stations are ready, Bob
initiates the transfer by sending a request code. Alice replies with an acknowledgment to indicate that she is also ready
to proceed to QKD. After the acknowledgment is received, Bob begins transmission of the synchronization sequence. For each
bit received Alice replies with a quantum state, modulated according with the PHM\_A value. She also stores the received
synchronization bit into SYNC\_A. Bob uses PHM\_B data to change the state of the receiving interferometer and stores
single-photon detector clicks into SPD\_B.

After the packet is transferred Alice copies SYNC\_A buffer to her laptop, and Bob does the same with the SPD\_B buffer.
The rest is performed using the laptops with TCP/IP connection between them. First, Alice and Bob compare the contents
of their synchronization buffers: SYNC\_A and SYNC\_B. If they differ, the packet is discarded and is not used as a raw
key. If they are the same, the key sifting is performed. Bob tells Alice the positions in SPD\_B when his detector
produced clicks. Alice creates her raw key from her PHM\_A data at the specified positions. Bob uses PHM\_B for the same
purpose, but inverts all the data. Ideally, Alice and Bob should arrive to the same key. However, experimental
imperfections and dark detector counts lead to errors. For the purpose of current publications the raw keys were
directly compared to calculate corresponding QBER and estimate the asymptotic secret key rate.

\subsection*{Relativistic protocol and the presence of air in the channel}

An important question is whether the protocol remain secure under the presence of air in the quantum channel. So
far in the model we assumed that all signals in the quantum channel propagate with the (vacuum) speed of light, which is not the
case for terrestrial line-of-sight atmospheric links. The answer directly depends on the channel length and the delay
$\Delta T$ between the two WCPs in the quantum channel. If $\Delta T$ cannot be compensated during the round trip by the eavesdropper
substituting a vacuum channel instead of the atmospheric one, the system remains perfectly secure, as all the assumptions
remain correct. Theoretically, increasing $\Delta T$ we can achieve secure operation even in the case of an optical
fiber based link. However, this becomes largely impractical as the required delay equals a significant fraction of the
communication distance. So the receiving delay interferometer needs to be almost as large and lossy as the channel
itself, which is undesirable. Going to the extreme, any channel type can be supported if one can guarantee that the
first WCP reaches Bob's setup {\em before} the second one leaves Alice's. This situation has much in common
with~\cite{GV95}, where this sequential quantum transfer was proposed for the first time, but with single photons.

For calculations we assume that the air refraction index is 1.0002804 that corresponds to the group velocity in dry air at
15~\textdegree C, 101.325~kPa and with 450~ppm CO$_2$ content at the wavelength of 780~nm.
The maximal channel length is given by
$$
	L_{\mathrm max} = \frac12\frac{c\Delta T}{n-1},
$$
which gives $L_{\mathrm max} = 10.7$~km at $\Delta T = 20$~ns. Therefore, current experimental realization is well within
the maximum range limited by the presence of air, so it is as secure as it would be with the vacuum quantum channel.

\subsection*{Receiving side interferometer alignment}

In order to operate properly, the receiving side delay interferometer must be aligned such that without any phase shifts
no light propagates into the single-photon detector (SPD). In practice, the required phase shift is constantly changing
due to thermal variations of the optical path lengths and also due to slow wavelength drifts. Typical time-scale of
these variations is in the order of a minute or even shorter.

\begin{figure}
\centering
\includegraphics[width=\columnwidth]{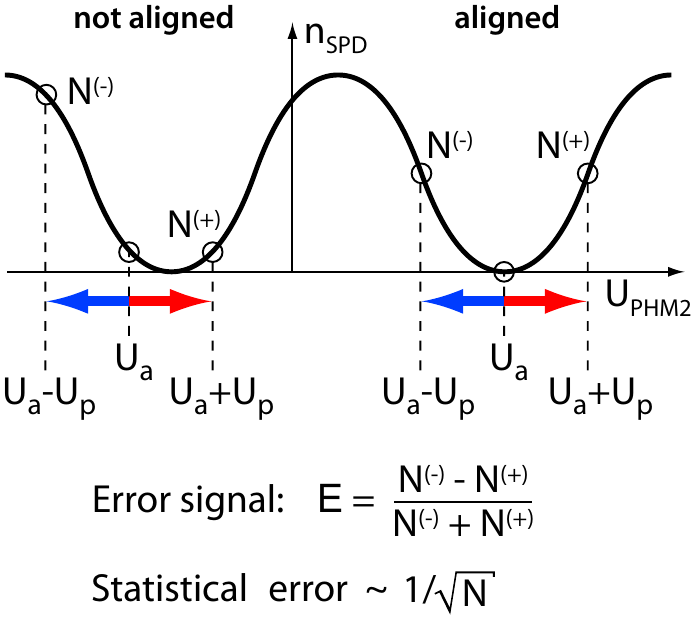}
{\caption{Receiving side interferometer transmission vs. phase modulator bias voltage. If the interferometer is not
properly aligned (left side), equal bias voltage offsets produce unequal numbers of detector clicks, which contribute to the error
signal correcting $U_a$. If it is already aligned (right side) error signal will be zero except for the statistical
deviations due to the limited number of detector clicks, which are averaged out with time. No $U_a$ correction is carried out in this
case.}\label{fig_alignment}}
\end{figure}

To solve the problem we implemented a closed loop control that takes the error signal from the SPD measurements and
adjusts the bias voltage. The error signal is a normalized difference between the number of detector counts when biased
above and below the current estimate, see Fig.~\ref{fig_alignment}.
The smaller the number of counts the larger are the statistical fluctuations of the error signal, so the obtained error
signal is scaled appropriately to ensure stable convergence to the best estimate.

The system uses two 4~ms long time windows to calculate detector counts when biased below and above the current value.
The rest of the 16~ms time frame is used for quantum key distribution.
Then all the steps are repeated again. The effective feedback speed depends on the signal strength and usually is at
least several Hz, which is enough to track the phase changes in real time.

The phase modulator can produce phase shifts of several wavelengths, but nevertheless sometimes the bias voltage needs to keep
increasing even when it hits its allowed maximum value. In this case our digital feedback scheme makes a step back by
several full wavelengths, decreasing the required voltage but keeping the same phase relations. The described scheme
proved to work well under broad range of conditions, and can be easily relied upon in practice.

\subsection*{Single mode free-space channel and tracking system}

The single mode free-space channel itself is an advanced piece of equipment. The presence of moving air in the channel makes the
link to have dynamic behavior and, therefore, requires an active tracking system.

The active tracking system corrects only for the most critical channel disturbance --- deviation of the beam from a
straight line. Higher order perturbations, not accounted in our approach, lead to beam profile distortion, which also
contributes to the coupling loss, but requires much more advanced adaptive optics tools to correct. Previous
reports~\cite{A07,A12} show satisfactory behavior of similar tracking systems at operation distances up to 1~km
at telecommunication wavelengths. This gives us confidence that such systems can be extended to even longer metropolitan
scale free-space single mode links, that makes our {\em relativistic} QKD approach viable and ready to substitute some
conventional fiber-based QKD methods. Another experiment~\cite{YSJ07} confirms that even a 150~km range is feasible
for a single mode free-space link. As the security of the proposed QKD protocol is totally decoupled from the channel
loss, future development of low dark count nanowire single photon detectors may overcome current system loss limitation
and make it suitable for 100~km range free-space QKD.

Operation of our tracking system does not require any classical communication channel between the stations. Each station
performs the following task: it measures the direction of the beacon light arrival and points the transmission beam
exactly at the same direction. Due to the link reciprocity, this makes sure that the transmitted beam always reaches the
destination. In other words, each station has its own closed loop control of the steering mirror, and this is enough for
a stable operation of the whole link.

Beacon light is slightly defocused such that it creates a half meter diameter spot at the end of the channel. Initial
setup requires coarse pointing of the stations to each other, so each of them could see the beacon light from the other.
After this is achieved, stations fall into the closed loop control mode and the quantum link becomes ready for QKD.
Importantly, after link downing when something blocks the beams, it reliably recovers by itself and does not need any
operator intervention.

\end{document}